\documentclass[aps,prd,twocolumn,superscriptaddress,amsmath, nofootinbib]{revtex4-1}

\usepackage[hidelinks]{hyperref}
\usepackage{mathtools}
\usepackage{aas_macros}
\usepackage[capitalize]{cleveref}

\Crefname{equation}{Equation}{Equations}
\crefname{equation}{Eq.}{Eqs.}
\Crefname{figure}{Figure}{Figures}
\crefname{figure}{Fig.}{Figs.}
\Crefname{table}{Table}{Tables}
\crefname{table}{Tab.}{Tabs.}
\Crefname{section}{Section}{Sections}
\crefname{section}{Sec.}{Secs.}

\newcommand{\p}{\mathrm{p}}
\newcommand{\m}{m_\p}
\newcommand{\mm}{m_\p^2}
\newcommand{\prm}[1]{{#1}^{\prime}}
\newcommand{\dprm}[1]{{#1}^{\prime\prime}}
\newcommand{\sft}{\beta}
\renewcommand{\d}[2][]{\operatorname{d}^{#1}\!{#2}}

\begin{document}

\title{Logolinear series expansions with applications to primordial cosmology}

\author{Will Handley}
\email[]{wh260@mrao.cam.ac.uk}
\affiliation{Astrophysics Group, Cavendish Laboratory, J.J.Thomson Avenue, Cambridge, CB3 0HE, UK}
\affiliation{Kavli Institute for Cosmology, Madingley Road, Cambridge, CB3 0HA, UK}
\affiliation{Gonville \& Caius College, Trinity Street, Cambridge, CB2 1TA, UK}

\author{Anthony Lasenby}
\email[]{a.n.lasenby@mrao.cam.ac.uk}
\affiliation{Astrophysics Group, Cavendish Laboratory, J.J.Thomson Avenue, Cambridge, CB3 0HE, UK}
\affiliation{Kavli Institute for Cosmology, Madingley Road, Cambridge, CB3 0HA, UK}

\author{Mike Hobson}
\email[]{mph@mrao.cam.ac.uk}
\affiliation{Astrophysics Group, Cavendish Laboratory, J.J.Thomson Avenue, Cambridge, CB3 0HE, UK}

\date{\today}

\begin{abstract}
    We develop a method for computing series expansions for solutions to ordinary differential equations when the asymptotic form contains both linear and logarithmic terms. Such situations are common in primordial cosmology when considering series expansions out of a singularity in the equations arising from a pre-inflationary phase of the universe. We develop mathematical techniques for generating these series expansions, and apply them to polynomial and Starobinsky inflationary potentials with kinetic initial conditions. Code for analytic and numerical computation of logolinear series is provided on GitHub.
\end{abstract}

\maketitle

\section{Introduction}\label{sec:introduction}

Even in the latest cosmic microwave background data~\cite{planck_inflation, planck_inflation2018,pip,sptpol}, tensions still exist at low multipoles with unexpected features and a suppression of power on large spatial scales both visible by eye in the $C_\ell$ spectra~\cite{WMAP1,WMAP2,WMAP3,Features}. 
Beyond the possibility that these effects are due just to chance, there are numerous possible physical explanations for these small anomalies~\cite{cmb_anomalies}, including inflationary models with singularities and discontinuities~\cite{H1,H2,H3}, multi-field phase-transitions~\cite{H4,H5,H6,H7,H8}, M-theory~\cite{H9,H10}, supergravity~\cite{H11}, just-enough inflation models~\cite{just_enough_inflation} or kinetic dominance~\cite{kinetic_dominance,Hergt1,Hergt2}. In this paper, we focus on the last of these.

It has been shown in general~\cite{kinetic_dominance} that classical solutions to the cosmic evolution equations begin in a big-bang singularity with the kinetic energy of the inflaton dominating over its potential energy. For this phase of kinetic dominance the solutions to the background equations take a generic and asymptotically simple form. After this phase of early kinetic dominance, evolution settles into its traditional slow-roll form. There is some evidence for kinetic dominance in current data~\cite{transients_finite,Hergt2}, although issues on how to set quantum initial conditions in such a phase, and the nature of their imprint on the cosmic microwave background are still under theoretical investigation~\cite{danielsson,quantum_initial_conditions}.
More importantly, there is more than enough wiggle-room in the currently observed primordial power spectrum of curvature perturbations to allow for these extensions, as well as the capacity for future experiments to constrain them~\cite{core_inflation}.

The asymptotic solutions examined in~\cite{kinetic_dominance} are merely the first terms in a series expansion. To compute higher order terms in the kinetic dominance approximation, we must consider a more general type of power series, which we term logolinear expansions. In this paper we generalise the results in~\cite{lasenby_doran} from the closed case to open, closed and flat cases, and from a quadratic potential to a field with general $V(\phi)$. We then apply this methodology to some example potentials.
These higher order terms may prove useful for improving the stability of numerical codes, for example those computing mode functions of primordial power spectra~\cite{mukhanov,ward}. It should be noted that whilst it remains true that flat universes are most preferred by current observational data~\cite{planck_parameters}, late-time curvature is only weakly coupled to primordial curvature, and thus does not preclude effects at inflation which could be of interest. 

The structure of this paper is as follows: In \cref{sec:background} we review the critical equations of inflationary cosmology to which we will apply our series expansion methodology. \cref{sec:logolinear} establishes notation and identities for logolinear expansions. In \cref{sec:methodology} we apply logolinear expansions to the equations of the early universe. \cref{sec:examples,sec:numerical_considerations} provide concrete applications to specific potentials and we conclude in \cref{sec:conclusion}.

\section{Background}\label{sec:background}

The background equations for a homogeneous Friedmann-Robertson-Walker spacetime with material content defined by a scalar field are:
\begin{align}
    \dot{H} + H^2 &= -\frac{1}{3\mm}\left( \dot{\phi}^2 - V(\phi) \right),
    \label{eqn:raychaudhuri}\\
    0&=\ddot{\phi} + 3 H\dot{\phi} + \frac{\d{}}{\d{\phi}}V(\phi),
    \label{eqn:klein_gordon}
\end{align}
where $H=\frac{d}{dt}\log a$ is the Hubble parameter, $\phi$ is the homogeneous value of the scalar field, $V(\phi)$ is the scalar potential, $a$ is the scale factor and dots indicate derivatives with respect to cosmic time $t$. One may supplement \cref{eqn:raychaudhuri,eqn:klein_gordon} with a third non-independent equation:
\begin{equation}
    H^2  + \frac{K}{a^2} = \frac{1}{3\mm}\left( \frac{1}{2}\dot{\phi}^2 + V(\phi) \right),
    \label{eqn:friedmann}
\end{equation}
where $K\in\{+1,0,-1\}$ is the sign of the curvature of the universe. For the remainder of this paper we will set the Planck mass to unity $\m=1$, but note that one may reintroduce $\m$ at any time by replacing $\phi\to\phi/\m$, $V\to V/\mm$.

Under general conditions~\cite{kinetic_dominance} the solutions to \cref{eqn:raychaudhuri,eqn:klein_gordon} begin in a singularity at $t=0$ with $\dot{\phi}^2\gg V(\phi)$. In this kinetically dominated regime, solutions take the asymptotic forms:
\begin{equation}
    \qquad H = \frac{1}{3t},\qquad \phi = \phi_\p \pm \sqrt{\frac{2}{3}}\log t,
    \label{eqn:kinetic_dominance}
\end{equation}
where $\phi_\p$ is a constant of integration.
The approximate solutions in \cref{eqn:kinetic_dominance} typically serve to set initial conditions for the numerical solution of differential \cref{eqn:raychaudhuri,eqn:klein_gordon}. Applying this prescription naively leads to difficulties, since \cref{eqn:friedmann} along with initial conditions in \cref{eqn:kinetic_dominance} effectively set the curvature to be slightly positive. This additional curvature can lead to problems with numerical stability where the (supposedly negligible) curvature terms can come to dominate before inflation begins, leading to collapsing solutions. More importantly, one expects \cref{eqn:raychaudhuri,eqn:klein_gordon} to encompass flat ($K=0)$, open ($K=-1$) and closed ($K=+1$) cases. To avoid confusion, note that the sign of $K$ is opposite in sense to the usually defined curvature density parameter $\Omega_K$, for example in a closed universe, $K=+1\Rightarrow \Omega_K<0$.

It is therefore natural to seek higher order terms to the kinetic dominance solutions in \cref{eqn:kinetic_dominance}. The singular nature of \cref{eqn:kinetic_dominance} as $t\to0$ renders a Taylor expansion inappropriate, as one requires negative powers of $t$ to describe $H$ terms. A more general Laurent expansion allowing for negative powers of $t$ will also not be applicable, as these do not include $\log t$ terms. By examining \cref{eqn:raychaudhuri,eqn:klein_gordon} one can see that expanding with \cref{eqn:kinetic_dominance} as leading order terms, one would need higher order terms such as $t\log t$. We are led therefore to consider the more general series expansions introduced by~\citet{lasenby_doran} which we term {\em logolinear expansions}.

\section{logolinear expansions}\label{sec:logolinear}
We will consider series expansions for a general function $x(t)$ in the form:
\begin{equation}
    x(t) = \sum_{j,k} [x^k_j] \: t^j {\left( \log t \right)}^k,
    \label{eqn:logolinear}
\end{equation}
where $[x^k_j]$ are twice-indexed real constants defining the series, with square brackets used to disambiguate powers from superscripts.
We make convenient definitions for upper and lower indexed functions of $t$ via partial summation:
\begin{gather}
    x_j(t)= \sum\limits_{k} [x_j^k]{(\log t)}^k, \qquad
    [x^k](t)= \sum\limits_{j} [x_j^k]\: t^{j}, \label{eqn:linlog_series} \\
    x(t) = \sum\limits_{j} x_j(t)\: t^{j} = \sum\limits_{k} [x^k](t)\: {(\log t)}^k.
\end{gather}
In this paper we are deliberately lax with bounds on the limits, as in general the bounds of $k$ depend on $j$ in a non-trivial way, and $j$ will be required to range over non-integer values. This approach was heavily influenced by \citet{concrete_mathematics}.

Care must be taken with logolinear expansions, as in general they are underdetermined. Writing out the first few terms of \cref{eqn:logolinear} for clarity:
\begin{align}
    x(t) = 
    & [x_0^0] &+& [x_1^0]\: t &+& [x_2^0]  t^2 
    &\ldots \nonumber\\
    & [x_0^1] \log t &+& [x_1^1]  t\log t  &+& [x_2^1]  t^2  \log t
    &\ldots \nonumber\\
    & [x_0^2] {(\log t)}^2 &+& [x_1^2]  t{(\log t)}^2  &+& [x_2^2]  t^2  {(\log t)}^2 
    &\ldots \nonumber\\
    &\vdots && \vdots && \vdots &\ddots,\label{eqn:explicit} 
\end{align}
and given that $t=\exp(\log t)$:
\begin{equation}
    t = 1 + \log t + \frac{1}{2!}{(\log t)}^2 +  \frac{1}{3!}{(\log t)}^3 + \ldots,
    \label{eqn:exp_log_t}
\end{equation}
then adding and subtracting $\alpha$ times \cref{eqn:exp_log_t} from \cref{eqn:explicit} yields:
\begin{align}
    x(t) = 
    & ([x_0^0]+\alpha)\: &+& ([x_1^0]-\alpha)\: t 
    &\ldots \nonumber\\
    & ([x_0^1]+\alpha)\: \log t &+& [x_1^1]\:  t\log t  
    &\ldots \nonumber\\
    & ([x_0^2]+\frac{\alpha}{2!})\: {(\log t)}^2 &+& [x_1^2]\:  t{(\log t)}^2  
    &\ldots \nonumber\\
    & ([x_0^3]+\frac{\alpha}{3!})\: {(\log t)}^3 &+& [x_1^3]\:  t{(\log t)}^3  
    &\ldots \nonumber\\
    &\vdots && \vdots &\ddots.
\end{align}
By setting $\alpha=[x_1^{0}]$ we can completely remove the $t$ term via a redefinition $[x_0^j] + \frac{1}{j!}[x_1^0] \rightarrow [x_0^j]$. Inspection shows that we can use a similar procedure to remove higher-order $t$ terms. Similarly, setting $\alpha=-[x_0^1]$, we may remove the $\log t$ term by redefining $[x_1^0]+[x_0^1]\to [x_1^0]$, $[x_0^j] - \frac{1}{j!}[x_0^1] \rightarrow [x_0^j]$, and again this may be performed for any term.

One may think of the underdetermination of logolinear series as a rather non-trivial gauge freedom which must be carefully controlled when using such series. In practice, we can avoid much of this difficulty by requiring the partial sums in \cref{eqn:linlog_series} over $\log t$ to truncate.

We may formally differentiate and integrate \cref{eqn:logolinear}:
\begin{gather}
    \dot{x} = \sum_{j,k} \left((j+1)[x_{j+1}^k] + (k+1) [x_{j+1}^k]\right) t^j {(\log t)}^k,
    \label{eqn:diff_ser}\\
    \int x \d{t} = \sum_{j>0,k} \left[\sum_{p\ge k}   \frac{-p![x^p_{j-1}]}{k!{(-j)}^{p-k+1}} \right] t^{j}{(\log t)}^k + c.
    \label{eqn:int_ser}
\end{gather}
Care must be taken in general with the limits of $j$ in \cref{eqn:diff_ser}, as negative $j$ indices have been introduced during the differentiation, and in deriving \cref{eqn:int_ser} we have used the identity:
\begin{equation}
    \int t^j {(\log t)}^k \d{t} = t^{j+1}\sum_{p=0}^k \frac{{-k!(\log t)}^p}{p!{(-(j+1))}^{k-p+1}} + c.
\end{equation}

Exponentials of the series in \cref{eqn:logolinear} may also be taken. We first define complete ordinary Bell polynomials~\cite{bell} via:
\begin{equation}
    \exp\left(\sum_{j} x_j t^j \right)
    = e^{x_0}\sum_{j} C_j(x_1,\dots,x_j)  t^j.
    \label{eqn:bell_polynomial}
\end{equation}
We may define and compute $C_j$ recursively via:
\begin{align}
     C_0 &=1,\nonumber\\
     C_j(x_1,\ldots,x_j) &= \sum_{k=1}^j \frac{k}{j} C_{j-k}(x_1,\ldots,x_{j-k}) x_k.
     \label{eqn:bell_recursion}
\end{align}
To derive the recursion in \cref{eqn:bell_recursion} from \cref{eqn:bell_polynomial}, take logarithms, differentiate with respect to and then multiply by $t$, multiply the consequent denominator and compare $t^j$ coefficients. The first few terms of \cref{eqn:bell_polynomial} are:
\begin{equation}
    e^{x_0}\left(1+ x_1 t + \left[ \frac{{x_1}^2}{2} + x_2 \right] t^2 + \left[ \frac{{x_1}^3}{6} + x_1 x_2 + x_3 \right]t^3 \right).\nonumber
\end{equation}
Note that for \cref{eqn:bell_recursion}, indices should range over integer values, but for \cref{eqn:bell_polynomial} $j$ is allowed to be non-integer. For more detail, see \citet{generatingfunctionology}. To exponentiate a logolinear series therefore, one simply applies the Bell polynomial expansion in \cref{eqn:bell_polynomial} to the lower indexed series expansion in \cref{eqn:linlog_series}. Care must be taken with the leading term, for example, if $x_0 = [x_0^0] + [x_0^1]\log t$ then
\begin{equation}
    e^{x_0} =  e^{[x_0^0] + [x_0^1]\log t} = e^{[x_0^0]}\cdot t^{[x_0^1]},
\end{equation}
we can see that exponentiating a logolinear series in this case will add a constant $x_0^1$-dependent shift to the powers of $j$ indices.
\section{Methodology}\label{sec:methodology}
We now apply logolinear series to the evolution equations of an inflating universe. We first examine the approach developed in~\cite{lasenby_doran}, before advocating a clearer approach that allows one to control gauge freedoms with more precision.

\subsection{The Lasenby-Doran approach (log-splitting)}\label{sec:log_splitting}
The approach espoused by \citet{lasenby_doran} is to substitute the partially summed expansions in \cref{eqn:linlog_series} ${H(t) = \sum_k [H^k](t){(\log t )}^k}$ and ${\phi(t) = \sum_k [\phi^k](t)  {(\log t )}^k}$ into master \cref{eqn:raychaudhuri,eqn:klein_gordon} to generate a set of recursion relations:
\begin{align}
    [H^{k+1}] =& \frac{t}{k+1}\left( \frac{1}{3}\left.V(\phi)\right|^k-[\dot{H}^k]  - \smashoperator{\sum_{p+q=k}}[H^p][H^q] + \frac{1}{3}[\phi^p][\phi^q]  \right),
    \nonumber\\
    [\phi^{k+2}] =& \frac{t^2}{(k+1)(k+2)}\Bigg([\phi^{k+1}]\frac{k+1}{t^2}- 2[\dot{\phi}^{k+1}]\frac{k+1}{t} - [\ddot{\phi}^k]
    \nonumber\\
    &+\left.\frac{\d{V}}{\d{\phi}}\right|^k-3\smashoperator{\sum_{p+q=k}}[H^p]\left[[\dot{\phi}^q] + [\phi^{q+1}]\frac{q+1}{t} \right] \Bigg).
    \label{eqn:recursion_ld}
\end{align}
When $[H^0]$, $[\phi^0]$ and $[\phi^1]$ are specified, all higher order functions $[H^k]$ may be calculated recursively.
Note that in \citet{lasenby_doran} only the $k=0$ case of recursion relations in \cref{eqn:recursion_ld} is explicitly stated.

We term this approach {\em log-splitting\/} since it involves substituting in partially summed series from \cref{eqn:linlog_series}, which are series in $\log t$ modulated by functions $x^k(t)$.

All that remain to be defined are initial power series $[H^0]$, $[\phi^0]$ and $[\phi^1]$, which should be: (a) power series in $t^{1/3}$ ``in order to generate curvature'' and (b) chosen so that ``successive terms in the series get progressively smaller''. The first of these statements can be seen by examining \cref{eqn:friedmann}, but the second is highly non-trivial, and intimately connected to the gauge freedoms indicated in \cref{sec:logolinear}. Indeed, the method presented in \citet{lasenby_doran} is only manually applied to the first few terms, as a log-splitting approach is extremely challenging to apply systematically.

\subsection{The lin-splitting approach}\label{sec:lin_splitting}
Given aforementioned issues with the approach outlined in the previous section, we now pursue an orthogonal methodology, which instead uses the lower-indexed partial sums from \cref{eqn:linlog_series}. This presents an easier way to systematically compute higher order terms whilst controlling gauge freedoms.

We begin by defining $N=\log a$, such that $\dot{N} = H$, so that \cref{eqn:raychaudhuri,eqn:klein_gordon} become:
\begin{align}
    \ddot{N} + \dot{N}^2 +\frac{1}{3}\left( \dot{\phi}^2 - V(\phi) \right) &=0,
    \label{eqn:raychaudhuri_N}\\
    \ddot{\phi} + 3 \dot{N}\dot{\phi} + \frac{\d{}}{\d{\phi}}V(\phi) &=0.
    \label{eqn:klein_gordon_N}
\end{align}
Using $N$ rather than $H$ puts \cref{eqn:raychaudhuri,eqn:klein_gordon} on a more equal footing, as it renders both evolution equations second order. 

Given that we intend to work with power series in $\log t$, we now transform \cref{eqn:raychaudhuri_N,eqn:klein_gordon_N} to logarithmic time, with $\prm{x}=\frac{\d{}}{\d{\log t}} x$:
\begin{align}
    \dprm{N} - \prm{N} + {\prm{N}}^2 +\frac{1}{3}\left( {\prm{\phi}}^2 - t^2V(\phi) \right) &=0,
    \label{eqn:raychaudhuri_x}\\
    \dprm{\phi} - \prm{\phi} + 3 \prm{N}\prm{\phi} + t^2\frac{\d{}}{\d{\phi}}V(\phi) &=0.
    \label{eqn:klein_gordon_x}
\end{align}
Converting to a first order system yields the definitions and equations:
\begin{align}
    \prm{N} &= h,
    \qquad 
    \prm{\phi} = v,
    \nonumber\\
    \prm{h} &= h - h^2 -\frac{1}{3} v^2 + \frac{1}{3} t^2V(\phi),
    \nonumber\\
    \prm{v} &= v - 3 v h - t^2\frac{\d{}}{\d{\phi}}V(\phi).
    \label{eqn:dsys}
\end{align}
We now substitute in our series definition from \cref{eqn:linlog_series}, and note that:
\begin{equation}
    x(t) = \sum_j x_j(t)\: t^j \quad\Rightarrow\quad \prm{x}(t) = \sum_j (\prm{x}_j + j x_j)\: t^j.
\end{equation}
We find that after equating coefficients of $t^j$, \cref{eqn:dsys} becomes:
\begin{align}
    \prm{N}_j + j N_j &= h_j,
    \qquad
    \prm{\phi}_j + j \phi_j = v_j,
    \nonumber\\
    \prm{h}_j + j h_j &= h_j + \frac{1}{3}\left. V(\phi)\right|_{j-2} -\smashoperator{\sum_{p+q=j}}h_p h_q+\frac{v_p v_q}{3},
    \nonumber\\
    \prm{v}_j + j v_j &= v_j - \left.\frac{\d{V(\phi)}}{\d{\phi}}\right|_{j-2} - 3 \smashoperator{\sum_{p+q=j}} v_p h_q.
    \label{eqn:dsysj}
\end{align}
It is also useful to consider the equivalent of \cref{eqn:friedmann}:
\begin{equation}
    \frac{1}{3}\left.V(\phi)\right|_{j-2} +\smashoperator{\sum_{p+q=j}}{\frac{1}{6}}v_p v_q - h_p h_q =\frac{K}{e^{2N_\p}}e^{\sum_{q>0}N_q(t)t^q}|_{j-\frac{4}{3}},
    \label{eqn:friedmann_logt}
\end{equation}
where exponentiation of logolinear series was discussed in \cref{sec:logolinear}.

For $j=0$, \cref{eqn:dsysj} is a non-linear differential equation in zero-indexed functions. Further inspection shows it to be equivalent to the starting equations~\cref{eqn:dsys} with $V=0$. Hence we may solve using the kinetically dominated solutions:
\begin{align}
    N_0 &= N_\p + \frac{1}{3}\log t, &h_0 &= \frac{1}{3}, \nonumber\\
    \phi_0 &= \phi_\p \pm \sqrt{\frac{2}{3}}\log t, &v_0 &=\pm\sqrt{\frac{2}{3}},
    \label{eqn:0_sol}
\end{align}
where $N_p$ and $\phi_p$ are constants of integration. Whilst we expect there to be four constants of integration {\em a-priori}, one of them is fixed by defining the singularity to be at $t=0$. As there are only two constants, it is clear that \cref{eqn:0_sol} does not span the full set of solutions to \cref{eqn:dsysj} with $j=0$. In fact, \cref{eqn:0_sol} represents a complete solution to $j=0$ for only the flat case, as $K=0$ effectively sets another integration constant. Nevertheless, we will discover that we may still use \cref{eqn:0_sol} as the base term for the logolinear series, and that the final constant of integration effectively emerges from a consideration of higher order terms.

For $j\ne 0$, we transfer terms involving $j$ in the summations from the right hand side to the left, giving a first order linear inhomogeneous vector differential equation:
\begin{equation}
    \prm{x}_j + A_j x_j = F_j,
    \label{eqn:linear_master}
\end{equation}
where $x=(N,\phi,h,v)$, $A_j$ is a (constant) matrix:
\begin{align}
    A_j &= \left(
    \begin{array}{cccc}
        j & 0 & -1 & 0 \\
        0 & j & 0 & -1 \\
        0 & 0 & j-1 + 2h_0 & \frac{2}{3}v_0 \\
        0 & 0 & 3v_0 & j-1+3h_0 \\
    \end{array}
    \right),\nonumber\\
    &= \left(%
    \begin{array}{cccc}
        j & 0 & -1 & 0 \\
        0 & j & 0 & -1 \\
        0 & 0 & j-\frac{1}{3} & \pm\frac{2}{9}\sqrt{6} \\
        0 & 0 & \pm\sqrt{6} & j \\
    \end{array}
    \right),\label{eqn:A}
\end{align}
and $F_j$ is a vector polynomial in $\log t$ depending only on earlier series $x_{p<j}$:
\begin{align}
    F_j &=
    \left(
    \begin{array}{c}
        0\\
        0\\
        \frac{1}{3}\left. V(\phi)\right|_{j-2}-\smashoperator{\sum_{\substack{p+q=j\\p\ne j,q\ne j}}}h_p h_{q} + \frac{1}{3}v_p v_q \\
        - \left.\frac{\d{V(\phi)}}{\d{\phi}}\right|_{j-2} - 3 \smashoperator{\sum_{\substack{p+q=j\\p\ne j,q\ne j}}} v_p h_q 
    \end{array}
    \right).\label{eqn:Fj}
\end{align}
Note that the limits on summations in \cref{eqn:Fj} now have strict bounds, in contrast with \cref{eqn:dsysj}. 

At each $j$, the linear differential \cref{eqn:linear_master} may be solved in terms of a complementary function with four free parameters and a particular integral. These free parameters correspond to the degrees of gauge freedom mentioned in \cref{sec:logolinear}.

\subsection{Complementary function}\label{sec:complementary_function}
We may solve the homogeneous version of \cref{eqn:linear_master} exactly, since $A_j$ is a constant matrix:
\begin{equation}
    \frac{\d{x_j^\mathrm{cf}}}{\d{\log t}} + A_j x_j^\mathrm{cf} = 0  \quad\Rightarrow\quad x_j^\mathrm{cf} = e^{-A_j\log t}[x_j^0],
\end{equation}
where $[x_j^0]$ is a constant vector parametrising initial conditions.
To compute the matrix exponential, we first compute eigenvectors and eigenvalues of $A_j$:
\begin{align}
    e_\sft &= \left(%
    \begin{array}{cccc}
        1& \pm\sqrt{6} & -1& \mp\sqrt{6} \\
    \end{array}
    \right),& 
    A_j e_\sft{} &= (j+1)\cdot e_\sft,
    \nonumber\\
    e_b &= \left(%
    \begin{array}{cccc}
        1& \mp\frac{3\sqrt{6}}{4}& \frac{4}{3}& \mp\sqrt{6}\\
    \end{array}
    \right),&
    A_j e_b &= (j-\frac{4}{3})\cdot e_b,
    \nonumber\\
    e_n &= \left(%
    \begin{array}{cccc}
        1& 0& 0& 0\\
    \end{array}
    \right),&
    A_j e_n &= j\cdot e_n,
    \nonumber\\
    e_\phi{} &= \left(%
    \begin{array}{cccc}
        0& 1& 0& 0\\
    \end{array}
    \right),&
    A_j e_\phi{} &= j\cdot e_\phi.\label{eqn:eigenvalues}
\end{align}
Parametrising initial conditions $[x_j^0]$ using the eigenbasis in \cref{eqn:eigenvalues} with parameters $\tilde{N},\tilde{\phi},b, \sft$, one finds:
\begin{align}
    x_j^\mathrm{cf}
    &= e^{-A_j\log t}(\tilde{N}e_n + \tilde{\phi}e_\phi -\tfrac{9}{14}b e_b + \sft e_\sft)\nonumber\\
    &= \left(\tilde{N}e_n + \tilde{\phi}e_\phi -\tfrac{9}{14} b e_b t^{4/3} + \sft e_b t^{-1}\right)t^{-j}.
    \label{eqn:complementary_function}
\end{align}

\Cref{eqn:complementary_function} is very interesting. First, all complementary functions (i.e.\ degrees of gauge freedom) are identical up to a $t^{-j}$ term, which cancels with $t^j$ in the power series in \cref{eqn:linlog_series}. We may set $\sft=0$ without loss of generality, as it grows faster than our leading term as $t\to0$. Choosing $\sft=0$ therefore amounts to setting the singularity to be at $t=0$ as an initial condition. We may absorb all $\tilde{N}$ and $\tilde{\phi}$ into our definitions of $N_\p$ and $\phi_\p$. The only remaining undetermined integration constant is $b$, which amounts to the third integration constant that was missing from \cref{eqn:0_sol}. The constant $b$ is controlled by the curvature of the universe via \cref{eqn:friedmann_logt}:
\begin{equation}
    K = b e^{2N_\p},
    \label{eqn:curvature_relation}
\end{equation}
where the somewhat cryptic $-\frac{9}{14}$ coefficient in \cref{eqn:complementary_function} was chosen so that the curvature relation in \cref{eqn:curvature_relation} takes a simple form.

Of particular importance is the fact that the curvature of the universe depends on a term in $t^{4/3}$. We should therefore be expanding as power series in $t^{2/3}$ rather than $t$. In our formalism, as espoused in~\cite{concrete_mathematics}, there is nothing preventing $j$ and its summation indices from being non-integer, so we happily do so.

\subsection{Particular integral}\label{sec:particular_integral}
All that remains to be determined is a particular integral of \cref{eqn:linear_master}, given that one has the form of $F_j$ at each stage of recursion.
The trial solution is $x_j(t)=\sum_{k=0}^{N_j} [x^k_j] {(\log t)}^k$. Defining ${F_j =\sum_{k=0}^{N_j} [F^k_j] {(\log t)}^k}$ and equating coefficients of ${(\log t)}^k$ yields
\begin{equation}
    (k+1)[x^{k+1}_j] + A_j [x^k_j] = [F^k_j],
    \label{eqn:linear_master_no_j}
\end{equation}
giving a descending recursion relation in $k$:
\begin{equation}
    [x^{N_j+1}_k]=0,\quad [x^{k-1}_j] = A_j^{-1}( [F_j^{k-1}]  - k [x^{k}_j]).
    \label{eqn:recursion_relation}
\end{equation}
The recursion relation in \cref{eqn:recursion_relation} fails when $A_j$ is non-invertible, which occurs when any of the eigenvalues in \cref{eqn:eigenvalues} are zero ($j=-1,0,4/3$). For these cases, the system is underdetermined, with an infinity of solutions parameterised along the directions of relevant eigenvectors. This infinity of solutions can therefore be carefully absorbed into a corresponding constant of integration.

Similarly, if we were to define an alternative base to the recursion in \cref{eqn:recursion_relation}, then infinite series would be generated. However, all but a finite number of terms would merely contribute to a re-definition of constants $N_\p$, $\phi_\p$, $b$, or an introduction of nonzero $\sft$, which we disallow due to the consequent shift of the singularity to a non-zero time $t$.

\section{Examples}\label{sec:examples}
We now apply these methods to two examples; first to that of a polynomial potential exemplified by a self-interacting field with a cosmological constant, and second to Starobinsky inflation as an example of a potential containing exponential terms.

The analytical calculations in this section were performed with the aid of the Maple${}^\mathrm{TM}$ (2017) computer algebra package~\cite{maple,Maple10}. The numerical calculations were performed using Python 3.6.4, for which the NumPy Polynomial package~\cite{scipy} was particularly useful. All code can be found on GitHub~\cite{github}.

\subsection{Polynomial potentials}\label{sec:polynomial}

Consider a non-trivial polynomial potential:
\begin{equation}
    V(\phi) = \mm \Lambda + \frac{1}{2}m^2 \phi^2 + \frac{1}{24}\lambda\phi^4,
\end{equation}
which incorporates both a cosmological constant $\Lambda$ and a self-interacting $\phi^4$ term. The application of the methods of \cref{sec:methodology} are straightforward, since we may extract the potential coefficients:
\begin{align}
    \left.V(\phi)\right|_j &= \mm\Lambda \delta_{0j} + \frac{1}{2}m^2\smashoperator{\sum_{p+q=j}}\phi_p\phi_q + \frac{1}{24}\lambda\smashoperator{\sum_{p+q+r+s=j}}\phi_p\phi_q\phi_r\phi_s,\nonumber\\
    \left.\frac{\d{V(\phi)}}{\d{\phi}}\right|_j &= m^2\phi_j + \frac{1}{6}\lambda\smashoperator{\sum_{p+q+r=j}}\phi_p\phi_q\phi_r,
    \label{eqn:pol_pot}
\end{align}
where $\delta_{ij}$ is the Kronecker delta function. As suggested in \cref{sec:complementary_function}, our logolinear series must be in powers of $t^{2/3}$, so that $j=0,\frac{2}{3},\frac{4}{3},2,\ldots$.
Substituting \cref{eqn:pol_pot} into the definition of $F_j$ in \cref{eqn:Fj}, we define $j=0$ solutions via \cref{eqn:0_sol}, $j=4/3$ solutions are computed using the complementary function as:
\begin{align}
    N_\frac{4}{3} &= -\frac{9}{14}b, &\phi_\frac{4}{3} &= \pm \frac{27\sqrt{6}}{56}b,\nonumber\\  h_\frac{4}{3} &= -\frac{6}{7}b, & v_\frac{4}{3} &= \pm\frac{9\sqrt{6}}{14}b,
    \label{eqn:pol_4_3}
\end{align}
and all remaining stages $\frac{2}{3},2,\frac{8}{3},\frac{10}{3},\ldots$ are computed from the recursion relation in \cref{eqn:recursion_relation}.
We find that the first few terms are:
\begin{widetext}
\begin{align}
H=&
\frac{1}{3t}-\frac{6}{7}b t^{1/3} 
\nonumber\\
&+ \Bigg[
\frac{\Lambda}{3}
+ \left( {\frac{4}{2187}}\mp{\frac {2  \sqrt {6}\phi_\p}{729}}+{\frac {\phi_\p^{2}}{81}}\mp{\frac {\sqrt {6}\phi_\p^{3}}{162}}+{\frac {\phi_\p^{4}}{72}} \right) \lambda
+ \left( {\frac{2}{81}}\mp\frac{\sqrt {6}\phi_\p}{27}+\frac{\phi_\p^{2}}{6} \right) {m}^{2}
\nonumber\\
&+ \left(  \left( -{\frac{4}{729}}\pm{\frac {2  \sqrt {6}\phi_\p}{243}}-\frac{\phi_\p^{2}}{27} \pm{\frac {\sqrt {6}\phi_\p^{3}}{54}} \right) \lambda+ \left( -{\frac{2}{27}}\mp\frac{\sqrt {6}\phi_\p}{9}   \right) {m}^{2} \right) \log t
\nonumber\\
&+ \left( \left( {\frac{2}{243}}\mp{\frac {\sqrt {6}\phi_\p}{81}}+\frac{\phi_\p^{2}}{18}  \right) \lambda+\frac{1}{9}{m}^{2}  \right) {(\log t)}^{2}
+ \left( -{\frac{2}{243}}\pm{\frac {\sqrt {6}\phi_\p}{81}} \right) \lambda {(\log t)}^{3}
+{\frac {\lambda }{162}{(\log t)}^{4}}
\Bigg]t
\nonumber\\
&-{\frac {702 {b}^{2}}{539}}{t}^{5/3}
\nonumber\\
&+ \Bigg[
{\frac {50 b\Lambda}{91}}
+ \left( {\frac {43210901 b}{1894708179}}\mp{\frac {4001293 b \sqrt {6}\phi_\p}{194329044}}+{\frac {152143 b\phi_\p^{2}}{3321864}} \mp{\frac {1507  \sqrt {6}b\phi_\p^{3}}{511056}}+{\frac {25 b\phi_\p^{4}}{1092}} \right) \lambda
\nonumber\\
&+ \left( {\frac {152143 b}{1660932}}\mp{\frac {1507 b \sqrt {6}\phi_\p}{85176}}+{\frac {25 b\phi_\p^{2}}{91}} \right) {m}^{2}
\nonumber\\
&+ \left(  \left( -{\frac {4001293 b}{97164522}}\pm{\frac {152143 b \sqrt {6}\phi_\p}{4982796}}-{\frac {1507 b\phi_\p^{2}}{85176}}\pm{\frac {25  \sqrt {6}b\phi_\p^{3}}{819}} \right) \lambda
+ \left( -{\frac {1507 b}{42588}}\pm{\frac {50 b \sqrt {6}\phi_\p}{273}} \right) {m}^{2} \right) \log t
\nonumber\\
&+ \left( \left( {\frac {152143 b}{4982796}}\mp{\frac {1507 b \sqrt {6} \phi_\p}{255528}}+{\frac {25 b\phi_\p^{2}}{273}} \right) \lambda+{\frac {50 b{m}^{2}}{273}} \right) {(\log t)}^{2}
\nonumber\\
&+ \left( -{\frac {1507 b}{383292}}\pm{\frac {50 b \sqrt {6}\phi_\p}{2457}} \right) \lambda {(\log t)}^{3}
 +{\frac {25 \lambda b}{2457}}{(\log t)}^{4}
\Bigg] {t}^{7/3},
\label{eqn:H_pol}
\end{align}
\begin{align}
\phi =&
\phi_\p \pm\sqrt{\frac{2}{3}}\log t
\pm{\frac {27 \sqrt {6}b }{56}}{t}^{4/3}
\nonumber\\
&+ \Bigg[ \mp\frac{\sqrt {6}\Lambda}{12}
+ \left( \pm{\frac {179  \sqrt {6}}{34992}}-{\frac {49 \phi_\p}{1944}}\pm{\frac {11  \sqrt {6}\phi_\p^{2}}{1296}}-{\frac {\phi_\p^{3}}{216}}\mp{\frac {\sqrt {6}\phi_\p^{4}}{288}} \right) \lambda
+ \left( \pm{\frac {11  \sqrt {6}}{648}}-\frac{\phi_\p}{36} \mp \frac{\sqrt {6}\phi_\p^{2} }{24}  \right) {m}^{2}
\nonumber\\
&+ \left(  \left( \mp{\frac {49  \sqrt {6}}{5832}}+{\frac {11 \phi_\p}{324}}\mp{\frac {\sqrt {6}\phi_\p^{2}}{216}}-\frac{\phi_\p^{3}}{36}  \right) \lambda+ \left( \mp{\frac {\sqrt {6}}{108}}-\frac{\phi_\p}{6} \right) {m}^{2} \right) \log t
\nonumber\\
&+ \left(  \left( \pm{\frac {11 \sqrt {6}}{1944}}-{\frac {\phi_\p}{108}}\mp{\frac {\sqrt {6}\phi_\p^{2}}{72}} \right) \lambda\mp \frac{\sqrt {6}}{36}{m}^{2}   \right) {(\log t)}^{2}
+ \left( \mp{\frac {\sqrt {6}}{972}}-{\frac {\phi_\p}{54}} \right) \lambda {(\log t)}^{3}
\mp{\frac {\sqrt {6}\lambda }{648}}{(\log t)}^{4}
\Bigg] {t}^{2}
\nonumber\\
&\pm{\frac {14337  \sqrt {6}{b}^{2}}{34496}}{t}^{8/3}
\nonumber\\
&+ \Bigg[ \mp{\frac {531  \sqrt {6}b\Lambda}{3640}}
+ \left( \mp{\frac {655319064583  \sqrt {6}b}{42104626200000}}+{\frac {3268631599 b\phi_\p}{35986860000}}\mp{\frac {13989847  \sqrt {6}b\phi_\p^{2}}{369096000}}+{\frac {27091 b\phi_\p^{3}}{946400}}\mp{\frac {177  \sqrt {6}b\phi_\p^{4}}{29120}} \right)\lambda
\nonumber\\
&+ \left( \mp{\frac {13989847  \sqrt {6}b}{184548000}}+{\frac {81273 b\phi_\p}{473200}}\mp{\frac {531  \sqrt {6}b\phi_\p^{2}}{7280}} \right) {m}^{2}
\nonumber\\
&+ \left(  \left( \pm{\frac {3268631599  \sqrt {6}b}{107960580000}}-{\frac {13989847 b\phi_\p}{92274000}}\pm{\frac {27091  \sqrt {6}b\phi_\p^{2}}{946400}}-{\frac {177 b\phi_\p^{3}}{3640}} \right) \lambda
+ \left( \pm{\frac {27091  \sqrt {6}b}{473200}}- {\frac {531 b\phi_\p}{1820}} \right) {m}^{2} \right) \log t
\nonumber\\
&+ \left(\left( \mp{\frac {13989847  \sqrt {6}b}{553644000}}+{\frac {27091 b \phi_\p}{473200}}\mp{\frac {177  \sqrt {6}b\phi_\p^{2}}{7280}} \right) \lambda
\mp{\frac {177  \sqrt {6}b{m}^{2}}{3640}} \right) {(\log t)}^{2}
\nonumber\\
&\pm \left( {\frac {27091  \sqrt {6}b}{4258800}}
-{\frac {59 b\phi_\p}{1820}} \right) \lambda {(\log t)}^{3}
\mp{\frac {59\lambda  \sqrt {6}b}{21840}{(\log t)}^{4}}
\Bigg] {t}^{10/3}.
\label{eqn:phi_pol}
\end{align}
\end{widetext}
The series in \cref{eqn:phi_pol,eqn:H_pol} above exhibit some comment-worthy properties. The curvature term $b$ does not begin mixing with potential terms $\lambda$, $m$, and $\phi_\p$ until the sixth term in each series. If one wishes to consider flat models therefore, a reasonable approach is to simply take $t$ and $t^2$ terms from the above as corrections to $H$ and $\phi$ as described in initial conditions in \cref{eqn:kinetic_dominance}, which effectively sets $b=K=0$ to first order.

\begin{figure}
    \centering
    \includegraphics{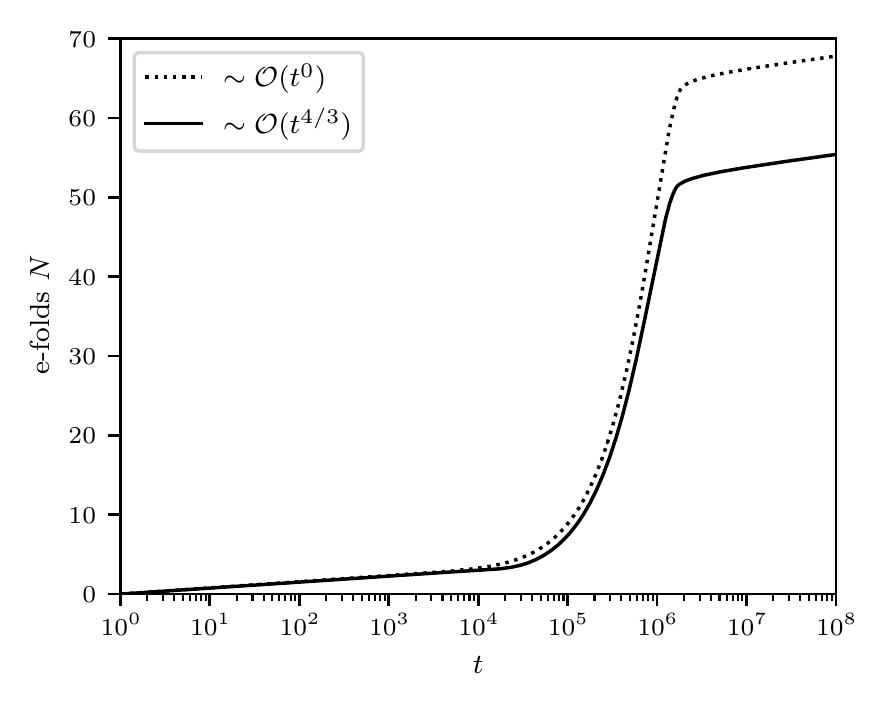}
    \caption{Inflation using $V=\frac{1}{2}m^2\phi^2$. We see generic features of kinetic initial conditions; the universe initially expands out of a singularity in kinetic dominance with $\dot\phi^2\gg V$, $a\propto t^{1/3}$. At later times, the potential comes to dominate $\dot\phi^2\ll V$, and a period of inflation begins at $t\sim m^{-1}$. Inflation then exits as the field $\phi$ oscillates about the minimum of the potential with the scale factor consequently expanding as some power law. For this example, we choose $m=10^{-5}\m$, consistent with current observations. As initial conditions at $t=1$ we set $\phi_\p=-23\m$, $N_\p=0$, in order to give $50$--$60$ e-folds of inflation, and choose a slightly closed universe $b=2\times10^{-6}$. Using only the first term in the kinetic dominance solution in order to start the numerical integration at $t=\m^{-1}$ gives the dotted line, whilst using the correction for curvature provided by the next term proves sufficient to match the true solution, which is indistinguishable from the solid curve.}\label{fig:N}
\end{figure}

\begin{figure}
    \centering
    \includegraphics{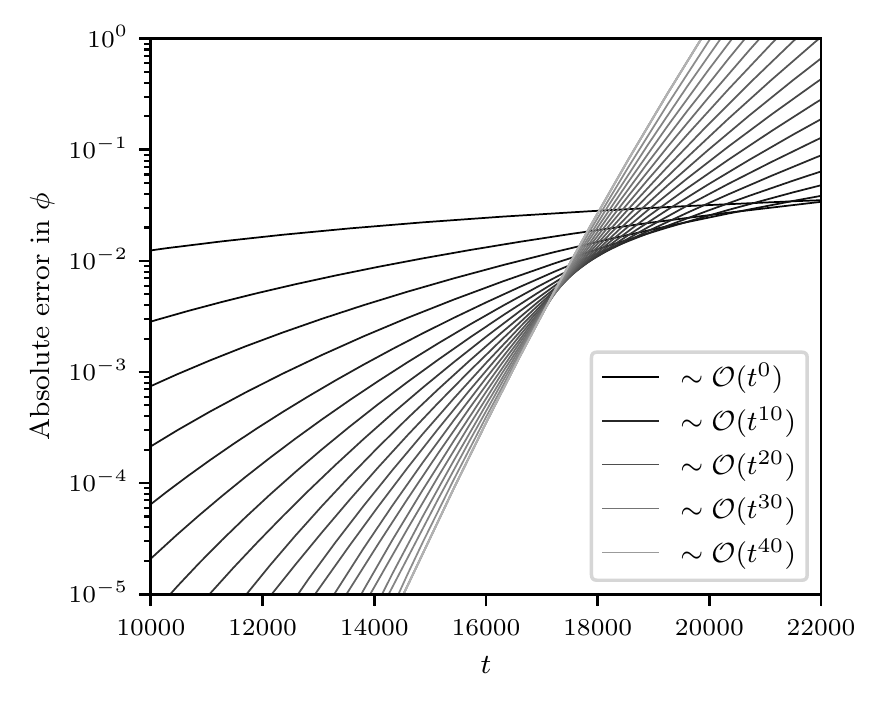}
    \caption{Absolute error in the expansion of $\phi$ for differing numbers of terms. Before $t\lesssim 17000\m$ the expansion becomes more accurate as the number of terms is increased, whereas for $t\gtrsim 17000\m$ the reverse is true. We use $V(\phi)=\frac{1}{2}m^2\phi^2$ with $m=10^{-5}\m$, $b=0$, $\phi_\p=-23\m$ on the positive branch and compare to a numerical solution defined by using the highest order approximation at $t=0$ as initial conditions. Plots for $H$ and $N$ are similar in quality.}\label{fig:phi_error}
\end{figure}

One can see the importance of including higher-order terms in \cref{fig:N}. For numerical codes aiming to provide constraints on the initial curvature of the universe, these corrections will be essential. \cref{fig:phi_error} shows the accuracy of the logolinear series as the number of terms is increased. In general, they are asymptotic series which provide an excellent approximation to the true solutions before inflation begins.

We may compare \cref{eqn:H_pol,eqn:phi_pol} with corresponding results from~\cite{lasenby_doran} by setting $\lambda=0$, reintroducing $\m$ via transformations $\phi\to\phi/\m$, $\phi_\p\to\phi_p/\m$, $m\to\mu/\m$, setting $\mm={(8\pi)}^{-1}$, changing variable definitions via $t=u$, $\phi_\p=b_0$, $b = -\frac{28\sqrt{6}}{81}\frac{b_4}{\m}$ and extracting relevant $\log t$ terms:
\begin{align}
    H^0=&
    \frac{1}{3u}+{\frac {32 \sqrt {3\pi}b_4 }{27}}{u}^{1/3} 
    -{\frac {6656  \pi {b_4}^{2}}{891}}  {u}^{5/3}
    \nonumber\\
    &+ \left( \frac{\Lambda}{3}+ {\frac{2\mu^2}{81}}\mp{\frac {4 \sqrt {3\pi}b_0\mu^2}{27}}+\frac{4\pi {b_0}^{2}\mu^2 }{3} \right) u, \nonumber\\
    \phi^0=&
    b_0\mp b_4 {u}^{4/3 } 
     \pm{\frac {118 \sqrt {3\pi}{b_4}^{2}}{99}} {u}^{8/3}
    \nonumber\\
    &+ \left( \mp{\frac {3\Lambda}{24\sqrt {3\pi}}}   \pm\frac {33 \mu^2}{1296 \sqrt {3\pi}}-\frac{b_0\mu^2}{36}\mp\frac{\sqrt {3\pi}{b_0}^{2}\mu^2}{6}  \right) {u}^{2}, \nonumber\\
    \phi^1=&
    \pm\frac {1}{\sqrt {12\pi}}+ \left( \mp{\frac {3\mu^2}{216 \sqrt {3\pi}}}-\frac{b_0\mu^2}{6} \right)  {u}^{2}.
    \label{eqn:lasenby}
\end{align}
\cref{eqn:lasenby} match precisely with results from~\cite{lasenby_doran}, up to $\pm$ branches, for which \citet{lasenby_doran} only consider the negative branch.

\subsection{Starobinsky inflation}\label{sec:starobinsky}

\begin{figure}
    \centering
    \includegraphics{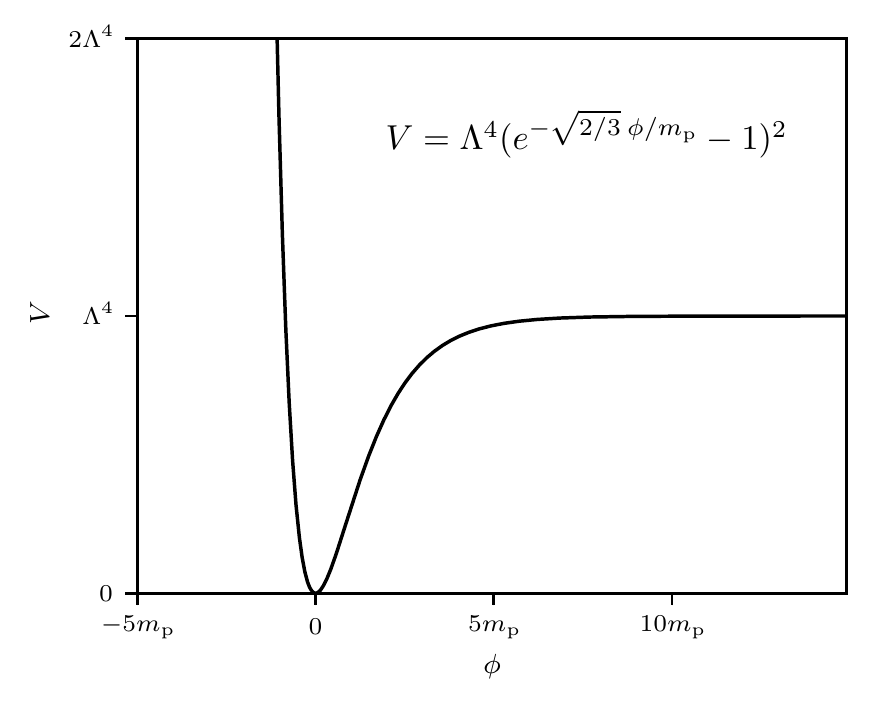}
    \caption{The Starobinsky potential for $R+R^2$ inflation in the Einstein frame. Its shape is typified by a plateau region at positive $\phi$, with an exponential slope at negative $\phi$.\label{fig:star}}
\end{figure}

Starobinsky~\cite{starobinsky} proposed a modified theory of gravity ($R+R^2$) as a mechanism for inflation, which remains one of the inflationary theories most consistent with observations~\cite{planck_inflation, planck_inflation2018}. After a conformal transformation to the Einstein frame the theory is equivalent to using a potential:
\begin{equation}
    V(\phi) = \Lambda^4{\left( e^{-\phi\sqrt{2/3}}-1 \right)}^2,
    \label{eqn:star_pot}
\end{equation}
which is plotted in \cref{fig:star}.
In light of discussions in \cref{sec:logolinear} of exponentiation of logolinear series, the factor of $\sqrt{2/3}$ in the Starobinsky potential is particularly convenient, as it allows us to keep indices as rational numbers (and is likely not a coincidence). 
If one were to consider more general exponential inflationary potentials, such as power law inflation~\cite{power_law_inflation} ($V\propto e^{-\lambda\phi}$), then we would be forced to introduce additional $t^q$ terms in the power series, where $q$ will not in general be a rational number.
 
From \cref{eqn:0_sol} we have $\phi_0 = \phi_\p \pm \sqrt{\frac{2}{3}}\log t$, and \cref{eqn:star_pot} becomes:
\begin{align}
    V(\phi) =& \Lambda^4 + 2\Lambda^4\Phi_\p e^{-\sqrt{2/3}\sum_{q\ne0}\phi_q t^q}t^{\mp2/3}  \nonumber\\
     &+\Lambda^4 \Phi_\p^2 e^{-2\sqrt{2/3}\sum_{q\ne0}\phi_q t^q}t^{\mp4/3},\label{eqn:star_terms}\\
     \Phi_\p =& e^{-\phi_\p\sqrt{2/3}}.\label{eqn:Phi_p_def}
\end{align}
We can see that exponential terms in general will shift the usual $j-2$ action of the potential terms in \cref{eqn:Fj} to $j-2\pm\frac{2}{3}$ and $j-2\pm\frac{4}{3}$ for the latter two terms in \cref{eqn:star_terms}.

Proceeding as for the polynomial case, $j=0$ terms are defined as in \cref{eqn:0_sol}, $j=\frac{2}{3}$ terms follow from the recursion relation in \cref{eqn:recursion_relation}. For $j=\frac{4}{3}$ terms, the negative branch is defined as for the polynomial case in \cref{eqn:pol_4_3}, but the positive branch must be handled with care, since the $\frac{2}{3}$ and $\frac{4}{3}$ terms now have contributions from the potential. In this case, we find:
\begin{align}
N_{\frac{4}{3}}^{+} &= -{\frac {9 b}{14}}-{\frac {9 {\Lambda}^{4}\Phi_\p}{14}}-{\frac {27 {\Lambda}^{8}\Phi_\p^{4}}{50}},
\nonumber\\
\phi_{\frac{4}{3}}^{+} &= {\frac {27 \sqrt {6}}{56} \left( b+\frac{2{\Lambda}^{4}\Phi_\p}{9} +{\frac {49 {\Lambda}^{8}\Phi_\p^{4}}{300}} \right) },
\nonumber\\
h_{\frac{4}{3}}^{+} &= -\frac{6}{7} b-\frac{6{\Lambda}^{4}\Phi_\p}{7} -{\frac {18 {\Lambda}^{8}\Phi_\p^{4}}{25}},
\nonumber\\
v_{\frac{4}{3}}^{+} &= {\frac {9 \sqrt {6}}{14} \left( b+\frac{2{\Lambda}^{4}\Phi_\p}{9} +{\frac {49 {\Lambda}^{8}\Phi_\p^{4}}{300}} \right) },
\end{align}
where the definition of $b$ has been judiciously chosen so that the curvature relation in \cref{eqn:curvature_relation} holds true. All remaining terms are defined via the recursion relation from \cref{eqn:recursion_relation}, and the first few terms for the two branches are:
\begin{widetext}
\begin{align}
H^{+}=&
\frac{1}{3t}+\frac{3\Phi_\p^{2}{\Lambda}^{4}}{5} t^{-1/3}
+ \left( -\frac{6}{7} b-3 \Phi_\p {\Lambda}^{4}-{\frac {18 \Phi_\p^{4}{\Lambda}^{8}}{25}} \right) t^{1/3}
+ \left(  \left( \frac{1}{3}+{\frac {111 b\Phi_\p^{2}}{70}} \right) {\Lambda}^{4}
+{\frac {38 \Phi_\p^{3}{\Lambda}^{8}}{35}}
+{\frac {819 \Phi_\p^{6}{\Lambda}^{12}}{1000}} \right) t
\nonumber\\
&+ \left( -{\frac {702 {b}^{2}}{539}}-{\frac {135 \Phi_\p b {\Lambda}^{4}}{98}}+ \left( -{\frac {1942 \Phi_\p^{2}}{2695}}-{\frac {10377 b\Phi_\p^{4}}{3850}} \right) {\Lambda}^{8}-{\frac {25023 \Phi_\p^{5}{\Lambda}^{12}}{15400}}-{\frac {52029 \Phi_\p^{8}{\Lambda}^{16}}{55000}} \right) {t}^{5/3}
\nonumber\\
&+ \Bigg[  \left( {\frac {50 b}{91}}+{\frac {539199 {b}^{2}\Phi_\p^{2}}{112112}} \right) {\Lambda}^{4}+ \left( {\frac {29 \Phi_\p}{91}}+{\frac {23169 b\Phi_\p^{3}}{5096}} \right) {\Lambda}^{8}+ \left( {\frac {470859 \Phi_\p^{4}}{280280}}+{\frac {535761 b\Phi_\p^{6}}{114400}} \right) {\Lambda}^{12}
\nonumber\\
&+{\frac {288081 \Phi_\p^{7}{\Lambda}^{16}}{114400}} +{\frac {133494399 \Phi_\p^{10}{\Lambda}^{20}}{114400000}} \Bigg] {t}^{7/3},
\label{eqn:H_plus_star}\\
H^{-} =&
\frac{1}{3t} -\frac{6b}{7} {t}^{1/3}+\frac{{\Lambda}^{4}}{3} t+ \left[ -{\frac {216 {b}^{2}}{539}}-{\frac {6 \Phi_\p {\Lambda}^{4}}{11}} \right] {t}^{5/3}+ \left[ {\frac {2 b}{91}}+\frac{3\Phi_\p^{2}}{13}  \right] {\Lambda}^{4}{t}^{7/3},
\label{eqn:H_minus_star}
\end{align}
\begin{align}
\phi^{+} =&
\phi_\p+\sqrt{\frac{2}{3}}\log t
+{\frac {3 {\Lambda}^{4}\sqrt {6}\Phi_\p^{2}}{20}}{t}^{2/3}
+ \left( {\frac {27 \sqrt {6}b}{56}}
+{\frac {3 \sqrt {6}\Phi_\p {\Lambda}^{4}}{28}}+{\frac {63 \sqrt {6}\Phi_\p^{4}{\Lambda}^{8}}{800}} \right) {t}^{4/3}
\nonumber\\
&+ \left(  \left( -\frac{\sqrt {6}}{12} -{\frac {33 \sqrt {6}b\Phi_\p^{2}}{35}} \right) {\Lambda}^{4}-{\frac {41 \sqrt {6}\Phi_\p^{3}{\Lambda}^{8}}{140}}-{\frac {441 \sqrt {6}\Phi_\p^{6}{\Lambda}^{12}}{2000}} \right) {t}^{2}
\nonumber\\
&+ \left( {\frac {14337 \sqrt {6}{b}^{2}}{34496}}+{\frac {891 \sqrt {6}\Phi_\p b{\Lambda}^{4}}{1568}}+ \left( {\frac {20047 \sqrt {6}\Phi_\p^{2}}{86240}}+{\frac {52263 \sqrt {6}b\Phi_\p^{4}}{35200}}\right) {\Lambda}^{8}+{\frac {19599 \sqrt {6}\Phi_\p^{5}{\Lambda}^{12}}{35200}}+{\frac {2386773 \sqrt {6}\Phi_\p^{8}{\Lambda}^{16}}{7040000}} \right) {t}^{8/3}
\nonumber\\
&+ \Bigg[  \left( -{\frac {531 \sqrt {6}b}{3640}}-{\frac {2863431 \sqrt {6}{b}^{2}\Phi_\p^{2}}{2242240}} \right) {\Lambda}^{4}+ \left( -{\frac {41 \sqrt {6}\Phi_\p}{455}}-{\frac {168813 \sqrt {6}b\Phi_\p^{3}}{101920}} \right) {\Lambda}^{8}
\nonumber\\
&+ \left( -{\frac {1549161 \sqrt {6}\Phi_\p^{4}}{2802800}}-{\frac {4855977 \sqrt {6}b\Phi_\p^{6}}{2288000}} \right) {\Lambda}^{12}-{\frac {2065941 \sqrt {6}\Phi_\p^{7}{\Lambda}^{16}}{2288000}}-{\frac{1025277831 \sqrt {6}\Phi_\p^{10}{\Lambda}^{20}}{2288000000}}
 \Bigg] {t}^{10/3},
\label{eqn:phi_plus_star}\\
\phi^{-} =&
\phi_\p-\sqrt{\frac{2}{3}}\log t+{\frac {27 \sqrt {6}b}{56}}{t}^{4/3}+ \frac{\sqrt {6}{\Lambda}^{4}}{12} {t}^{2}+ \left[ {\frac {6075 \sqrt {6}{b}^{2}}{34496}}-{\frac {15 \sqrt {6}\Phi_\p {\Lambda}^{4}}{88}} \right] {t}^{8/3}+ \left[ {\frac {21 \sqrt {6}\Phi_\p^{2}}{260}-{\frac {9 \sqrt {6}b}{520}}} \right] {\Lambda}^{4}{t}^{10/3}.
\label{eqn:phi_minus_star}
\end{align}
\end{widetext}
The most striking feature of \cref{eqn:H_plus_star,eqn:H_minus_star,eqn:phi_plus_star,eqn:phi_minus_star} is that there are no higher-order $\log t$ terms. This is somewhat to be expected, as under a variable transformation such as $\varphi = e^{\phi}$, the evolution \cref{eqn:raychaudhuri,eqn:klein_gordon} can be shown to no longer have any exponential terms. Since $\varphi$ is also power-law at early times, this means that one would not expect series expansions to require higher-order $\log t$ terms. It is reassuring that our methodology for logolinear series expansions is robust enough to recover this result without modification.

\begin{figure}
    \centering
    \includegraphics{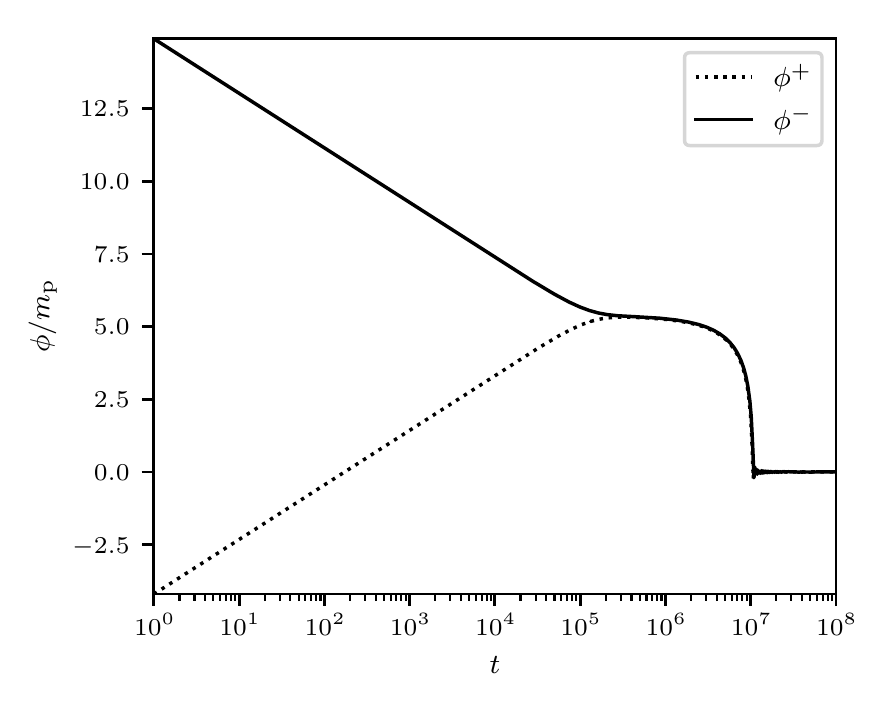}
    \caption{Evolution of the field $\phi$ in a Starobinsky potential. The two branches of the kinetically dominated solutions eventually come to rest in a slow roll inflating phase at $\phi\sim 5\m$, before exiting inflation and dropping into tight oscillations about $\phi=0$. Here we choose $\Lambda^2=10^{-5}\m$ for consistency with current observations, and $\phi_\p^+ = -4.2\m$, $\phi_\p^- = 14.9\m$, $b=0$, to give $50$--$60$ e-folds of inflation in both cases.}\label{fig:phi_star}
\end{figure}

For the positive branch of the Starobinsky solutions in \cref{eqn:H_plus_star,eqn:phi_plus_star}, unlike the polynomial case in \cref{eqn:H_pol,eqn:phi_pol}, we find that there are $t^{-1/3}$ and $t^{2/3}$ terms, and potential terms $\Phi_\p$, $\Lambda$ mix with curvature $b$ at lower order. At negative $\phi$ the Starobinsky potential grows faster than curvature as $t\to0$, in contrast with the polynomial case. For the negative branch of the Starobinsky solutions in \cref{eqn:H_minus_star,eqn:phi_minus_star}, in comparison with the polynomial case much higher order is required before potential terms become included. At positive $\phi$, the effect of the potential is weak at early times due to its plateau-like nature. Numerical solutions are plotted in \cref{fig:phi_star}. For the positive branch in particular, higher order terms are essential for numerical stability.

\section{Numerical considerations}\label{sec:numerical_considerations}
\vspace{-0.7em}

We now demonstrate how such series may be used to verify the numerical accuracy of kinetically dominated approximations. \citet{transients_finite} consider a $V=\frac{1}{2}m^2\phi^2$ potential with kinetic initial conditions. In their work, they parameterise the initial conditions in terms of a dominance factor $r=100$, so that at some early time:
\begin{align}
    \phi = \frac{\dot\phi}{r m} = \phi_0,\:\:  h=\frac{H}{a} = \sqrt{\frac{r^2+1}{6}} m\phi_0,\:\: a = a_0,
    \label{eqn:transient_conditions}
\end{align}
where $h$ now denotes the conformal Hubble factor. Substituting \cref{eqn:transient_conditions} into \cref{eqn:friedmann} shows that $h$ has been chosen to be consistent with a flat universe. Setting initial conditions in this way is common, as it does not require one to specify a time at which to set them. Nevertheless when \citet{transients_finite} go on to solve the Mukhanov-Sazaki equation, they implicitly assume that the initial conditions are set at a conformal time $\eta = \frac{a}{2h}$ after the start of the universe, where $\eta=\int_0^t \frac{dt}{a}$.
To generate series consistent with \citet{transients_finite} we take the negative branch and $\Lambda=\lambda=b=0$ in \cref{eqn:H_pol,eqn:phi_pol}, yielding:
\pagebreak

\begin{widetext}
\begin{align}
\phi =&
\phi_\p-\sqrt{\frac{2}{3}}\log t &+ \left[  \left( -{\frac {11 \sqrt {6}}{648}}-\frac{\phi_\p}{36}+\frac{\sqrt {6}{\phi_\p}^{2}}{24}  \right) {m}^{2}+ \left( {\frac {\sqrt {6}}{108}}-\frac{\phi_\p}{6} \right) {m}^{2}\log t+\frac{\sqrt {6}{m}^{2}}{36} {\log t}^{2} \right]& {t}^{2},
\label{eqn:phi_trans}\\
H =&
\frac{1}{3t} &+ \left[  \left( {\frac{2}{81}}+\frac{\sqrt {6}\phi_\p}{27} +\frac{{\phi_\p}^{2}}{6}  \right) {m}^{2}+ \left( -{\frac{2}{27}}-\frac{\sqrt {6}\phi_\p}{9}  \right) {m}^{2}\log t+\frac{{m}^{2}}{9} {(\log t)}^{2} \right]& t,
\label{eqn:H_trans}\\
a e^{-N_\p} =&
t^{1/3} &+ \left[  \left( {\frac{19}{324}}+{\frac {5 \sqrt {6}\phi_\p}{108}}+\frac{\phi_\p^{2}}{12} \right) {m}^{2}+ \left( -{\frac{5}{54}}-\frac{\sqrt {6}\phi_\p}{18}  \right) {m}^{2}\log t+\frac{{m}^{2}}{18} {(\log t)}^{2} \right]& {t}^{7/3},
\label{eqn:a_trans}\\
\eta e^{N_\p} =&
\frac{3}{2}{t}^{2/3} &+ \left[  \left( -{\frac{565}{13824}}-{\frac {29 \sqrt {6}\phi_\p}{1152}}-\frac{\phi_\p^{2}}{32}  \right) {m}^{2}+ \left( {\frac{29}{576}}+\frac{\sqrt {6}\phi_\p }{48} \right) {m}^{2}\log t-\frac{{m}^{2}}{48} {(\log t)}^{2} \right]& {t}^{8/3}.
\label{eqn:eta_trans}
\end{align}
\end{widetext}
We compute $a$ in \cref{eqn:a_trans} by exponentiating the series for $N$ as detailed in \cref{sec:logolinear}. The series in \cref{eqn:eta_trans} for $\eta$ is computed by performing a negative exponential on $N$ and then integrating the consequent series via \cref{eqn:int_ser}. These provide a set of relations which may be solved numerically to transform the conditions between $(\phi_0,a_0,r)$ in \cref{eqn:transient_conditions} and $(\phi_\p,N_\p,t)$. We may then use these to test the validity of the assumption that the conditions are effectively set at $\eta=\frac{a}{2h}$.

Taking $m=6\times10^{-6}$ as in \cref{eqn:transient_conditions}, a typical set of parameters transforms as:
\begin{align}
    (\phi_0, a_0, r) &= (20.7, 1, 100),\quad \frac{a}{2h} = 98.606 \nonumber\\ 
    \Rightarrow (\phi_\p,N_\p,t) &= (23.4176,-1.3952,65.7395) \nonumber\\
    \Rightarrow \eta &= 98.610.
\end{align}
We can therefore see that $\eta\approx \frac{a}{2h}$, accurate to within a fractional error of $10^{-4}$. \citet{transients_finite} have indeed set their initial conditions with sufficient accuracy, and having access to these power series makes for simple cross-checking of numerical stability.

\section{Conclusion}\label{sec:conclusion}
We developed techniques required for applying logolinear power series to the background differential equations of a Friedmann-Robertson-Walker universe, paying particular attention to details of how to control the gauge freedoms inherent in these expansions. We then applied our methodology to specific cases of polynomial and Starobinsky inflationary potentials, showing that our approach can be successfully applied to finite polynomial and exponential potentials. Future work will involve applying these series to a programme investigating conformally constrained closed universes.

Logolinear expansions could prove useful for improvement of stability of numerical integration codes requiring accurate background solutions such as those codes that solve for mode functions in the Mukhanov-Sazaki equation. The series we derive are particularly relevant to researchers aiming to observationally constrain just enough inflation models, as a lack of attention to higher order terms can lead to systematic errors in parameter constraints, particularly in the case of curvature.

\begin{acknowledgments}
    W.H.\ would like to thank Gonville~\&~Caius College for their ongoing support, and Robert Knighton \& Panagiotis Mavrogiannis for their preliminary investigations into these series solutions.
\end{acknowledgments}

\bibliographystyle{unsrtnat}
\bibliography{logolinear}

\end{document}